%% file: 0_main.tex
\documentclass[man]{apa7}

\usepackage[american]{babel}
\usepackage[style=apa,sortcites=true,sorting=nyt,backend=biber]{biblatex}
\DeclareLanguageMapping{american}{american-apa}
\usepackage{graphicx}
\usepackage{amsmath}
\usepackage{amsthm} 
\usepackage{amssymb}
\usepackage{booktabs}
\usepackage{ltablex}
\usepackage{longtable}
\usepackage{multirow}
\usepackage{setspace}
\usepackage{url}
\usepackage{subcaption}
\usepackage{tikz}
	\usetikzlibrary{arrows.meta,positioning} 
\usepackage{caption}
\usepackage{setspace}
\usepackage{csquotes}
\usepackage[symbol]{footmisc}

\AtEndPreamble{\hypersetup{hidelinks}}

\DeclareCaptionLabelSeparator*{spaced}{\\[2ex]}
\makeatletter
\renewcommand*{\@fnsymbol}[1]{\ensuremath{\ifcase#1\or \dagger \or \ddagger\or * \or
\mathsection\or \mathparagraph\or \|\or **\or \dagger\dagger
\or \ddagger\ddagger \else\@ctrerr\fi}}
\makeatother

\title{On the Nuisance of Control Variables in Causal Regression Analysis}
\shorttitle{On the Nuisance of Control Variables}

\authorsnames[1,2]{Paul H\"unermund,Beyers Louw}
\authorsaffiliations{{Copenhagen Business School, Department of Strategy and Innovation, Kilevej 14A, 2000 Frederiksberg, Denmark. Email: phu.si@cbs.dk},
{University of Groningen, Department of Innovation Management and Strategy, Nettelbosje 2, 9747 AE Groningen, The Netherlands. Email: beyers.louw@rug.nl}}

\leftheader{H\"unermund and Louw}

\abstract{Control variables are included in regression analyses to estimate the causal effect of a treatment on an outcome. In this paper, we argue that the estimated effect sizes of controls are unlikely to have a causal interpretation themselves, though. This is because even valid controls are possibly endogenous and represent a combination of several different causal mechanisms operating jointly on the outcome, which is hard to interpret theoretically. Therefore, we recommend refraining from interpreting marginal effects of controls and focusing on the main variables of interest, for which a plausible identification argument can be established. To prevent erroneous managerial or policy implications, coefficients of control variables should be clearly marked as not having a causal interpretation or omitted from regression tables altogether. Moreover, we advise against using control variable estimates for subsequent theory building and meta-analyses.}


\authornote{
  \addORCIDlink{Paul H\"{u}nermund:}{0000-0001-9163-038X}
  
  \addORCIDlink{Beyers Louw:}{0000-0001-8413-3959}
   
  This paper was published in \emph{Organizational Research Methods} under the DOI: \url{https://doi.org/10.1177/10944281231219274}.
  }

\begin{document}
\maketitle

\input{1_introduction}
\input{2_lit_review}
\input{3_theory}
\input{4_example}
\input{5_discussion}

\printbibliography

\appendix
\clearpage
\setcounter{page}{1}

\label{app:literature}
\section{Literature Review}
\label{appendixA}

The following tables list papers published in \emph{Strategic Management Journal} and \emph{Organization Science} between January 2015 and December 2020 that were identified to provide substantive interpretations of control variable estimates. Column ``Total" refers to the total number of papers in the respective journal issue employing parametric regression models and "Count" indicates the number of articles including sections that interpret controls. Example quotes are provided in Table \ref{tab:quotes}.
\input{tables/appendix_smj}

\input{tables/appendix_orgsci}

\section{Examples of control variable interpretation}
\label{appendixB}

Table \ref{tab:quotes} displays examples of how control variables have been interpreted in articles that were published in \emph{Organization Science} and \emph{Strategic Management Journal} between 2015 and 2020. We identified four broad categories of ways in which researchers interpret control variable estimates. 
\begin{enumerate}
    \item Mentioning that control variables behave as expected based on results found in prior literature
    \item Noting possibly interesting results regarding the sign and significance of control variable coefficients
    \item Taking control variable estimates as evidence that confirms results in prior literature
    \item Providing ex-post rationalizations of control variable estimates based on prior theory
\end{enumerate}

\input{tables/quotes}

\end{document}

%% file: 1_introduction.tex
Multivariate regression is an important tool for empirical research in organization studies, management, and economics. Beyond settings in which regression analysis is used to statistically predict a left-hand side variable given a set of explanatory variables, the main purposes of these methods is to control for confounding influence factors between a treatment and an outcome in order to obtain consistent causal effect estimates.\footnote{We refer to consistency in the statistical sense that the estimator converges in probability to the true underlying parameter \parencite[Def.\ 3.8]{Wooldridge2002}.} However, in practice scholars often overstate the role of control variables in regressions. In this paper we argue that, while essential for the identification of causal effects, control variables do not necessarily have a causal interpretation themselves. This is because even \emph{valid} controls are often correlated with other unobserved factors, which render their marginal effects uninterpretable from a causal inference perspective \parencite{Westreich2013, Keele2020}. Consequently, researchers need to be careful with attaching too much meaning to control variables and should consider to ignore them when interpreting the results of their analyses.

Drawing substantive conclusions from control variable estimates is common, however. Authors frequently make use of formulations such as: ``control variables have expected signs" or ``it is worth noting the coefficients\footnote{In linear regression models, marginal effects of explanatory variables are constant and can be represented by a single parameter. In the following we will therefore mainly refer to regression coefficients of control variables although our arguments also apply to marginal effect estimates obtained from nonlinear models.} of our control variables". Below, we present the results of a literature review of papers published in \emph{Organization Science} and \emph{Strategic Management Journal} between 2015 and 2020, in which we found that 47 percent of manuscripts using regression methods also explicitly discussed the estimated effect sizes of controls. This is in line with \Textcite{Carlson2012}, who identified 48 percent of papers published in the Academy of Management Journal, Journal of Applied Psychology and Strategic Management Journal in 2007 that interpreted and discussed the effects of controls. Moreover, in our own experience as authors of quantitative research papers, we frequently encountered instances in which reviewers asked us to provide an interpretation of control variable coefficients. The justification that was often given was that, although they were not the main focus of the analysis, controls could still provide valuable information for other researchers in the field who are investigating related research questions. 

The methodological literature in organizational research usually highlights that control variables should carry the same importance in an empirical analysis as the main independent variables of interest \parencite{Becker2005, Spector2011, Carlson2012, Atinc2012}. To increase rigor and improve transparency of published research articles, \Textcite{Becker2005} recommends to report all regression coefficients of control variables as well as their significance levels. Similarly, \Textcite{Spector2011} advocate that controls should be given equal status to the main treatment variable in the analysis. \Textcite{Atinc2012} consider it to be best practice to provide an ex-ante prediction of the sign of the relationship between the controls and outcome variable based on theory, which should subsequently be checked against the empirical evidence. In a recent paper, \Textcite{becker2016} provide a more cautionary recommendation regarding generalizing from control variable estimates if it involves out-of-sample extrapolation, but otherwise considers it to be appropriate. Overall, the general consensus in the organizational literature thus seems to be that interpreting control variable estimates is safe, as it adds to the body of cumulative evidence regarding a particular effect size. 

This paper builds on the graphical framework to causality \parencite{Pearl2000, Durand2009, Huenermund2023}. Causal diagrams have already been established as a powerful tool for determining which control variables are relevant to a given regression model \parencite{Cinelli2022, Huenermund2022}. In addition, the approach offers a distinct perspective on the proper interpretation and communication of control variable results, which differs from prior methodological practices found in the organizational and management literature. In the following, we will explicate the view that control variables, while certainly an important ingredient in many causal research designs, do not have the same status as the main variables of interest in an empirical analysis. In particular, we will argue that in many situations valid controls can nonetheless be \emph{endogenous}. Therefore, interpreting their estimated effect sizes in light of prior theory could lead to potentially misleading conclusions. A valid causal interpretation of control variables rests on strong assumptions and usually requires accounting for all influence factors of the outcome variable under study. Since this is unlikely to be fulfilled in many research contexts, we recommend authors to exercise caution when interpreting control variables and consider omitting estimated coefficients of control variables from regression tables, or relegating them to an appendix. Finally, we discuss what our recommendations imply for the practice of meta-analysis, which has recently gained traction in many fields including organizational research \parencite{Anguinis2011}.

%% file: 2_lit_review.tex
\section{Do researchers attach substantive meaning to control variables?}

To assess the degree to which researchers interpret control variable estimates in their studies, we conducted a review of all articles published in \emph{Organization Science} and \emph{Strategic Management Journal} between January 2015 and December 2020. We chose these two journals because of their high prestige in the management and organization field as well as their reputation for high-quality empirical research. Our sample includes all quantitative articles that employed parametric regression models such as OLS, logit, probit, Poisson, etc. This choice was made because effect sizes of control variables can usually not be summarized by a single coefficient (or marginal effect) in non- and semiparametric models. The use of such methods is rare in our sample anyway though. 

We manually categorized papers according to whether they interpret or draw substantive insights from the coefficients or marginal effect estimates of control variables. Examples of such an interpretation range from ``the control variable CEO tenure is positively related to performance" to ``the effect sizes of of control variables are in line with previous studies". The latter interpretation is thereby of relevance because authors of future research papers might be tempted to develop theory based on this seemingly accumulating empirical evidence. The result of our review shows that interpreting control variables was common practice in the analyzed journals during our period of observation. For the Strategic Management Journal, we identified a total of 497 quantitative research articles, of which 233 (47 percent) proceeded to interpret the effects of controls variables. For Organization Science, out of a total of 275 quantitative articles, 131 (47 percent) provided an interpretation of control variable estimates. Detailed results of the literature review and example quotes demonstrating the different ways in which authors interpret control variable estimates in practice are reported in the supplemental material to this paper.

%% file: 3_theory.tex
\section{The causal interpretation of control variables}

The relationship between the main explanatory variables and controls in a regression can be complex, therefore it is beneficial to explicitly depict them in a causal diagram \parencite{Pearl2000}. \Textcite{Durand2009} were the first to introduce causal graphs as a tool to management researchers. \Textcite{Cinelli2022} provide a useful overview of the different functions of control variables in regression analyses, explicitly leveraging the graphical framework. Here, we focus on the causal interpretation and reporting of control variable estimates, which has been a topic of ongoing debate in the organizational literature.

Figure \ref{fig:1a} presents a simple model with a treatment variable $X$ and an outcome variable $Y$. Both variables are connected by an arrow, denoting the direction of causal influence between them. In addition, there are two confounding variables, $Z_1$ and $Z_2$, that are affecting the treatment and the outcome. $Z_1$ and $Z_2$ are correlated, as a result of a common influence factor they share, which is denoted by the dashed bidirected arc in the graph. The fact that $Z_1$ and $Z_2$ are correlated creates what is known as a backdoor path between the treatment and the outcome \parencite{Pearl2000}. $X$ and $Y$ are not only connected by the direct causal path $X \rightarrow Y$, but also by a second path, $X \leftarrow Z_1 \dashleftarrow\dashrightarrow Z_2 \rightarrow Y$, which creates a spurious, non-causal correlation between them. 

\hspace{1pt}
\centerline{
------
Insert Figure \ref{fig:1} about here
------
}
\hspace{1pt}

\input{figures/figure1}

Backdoor paths are defined as any sequence of arrows connecting the treatment and outcome variable (irrespective of their orientation) that remains if arrows emitted by the treatment are deleted from the graph \parencite{Pearl2000}. Because of the latter requirement they are easy to spot in the causal diagram. Since all the arrows emitted by $X$ are deleted, backdoor paths have to point into $X$ instead; i.e., they enter ``through the backdoor", which is where the name comes from.

Control variables in a multivariate regression model are invoked to block such backdoor paths and obtain a consistent estimate of the causal effect of $X$ on $Y$, in which case one speaks of an effect to be \emph{causally identified}. For this purpose, it is sufficient to control for any variable that lies on the open path.\footnote{
Technical note: Requiring the path to be previously unblocked rules out that the variable which is adjusted for is a collider \parencite{Huenermund2023}. A discussion of colliders and bad controls goes beyond the scope of this note \parencite[see][for a detailed discussion.] {Cinelli2022, Huenermund2022}.}
Thus, in the example of Figure \ref{fig:1a}, the researcher has the choice between either controlling for $Z_1$ or $Z_2$, since both would allow to identify the causal effect of interest. The choice between different admissible sets of control variables is thereby of high practical relevance. Researchers often have fairly detailed knowledge about the treatment assignment mechanism $Z_1 \rightarrow X$; e.g., because there are organizational or administrative rules that determine individual treatment status, which can be exploited for identification purposes \parencite{Angrist1990,Flammer2017}. At the same time, the set of variables $Z_2$ that are direct influence factors of $Y$ will likely be large. Thus, in practical applications it might be much easier to control the treatment assignment mechanism instead of trying to include all variables that have an effect on the outcome in a regression.

\hspace{1pt}
\centerline{
------
Insert Table \ref{tab:sim_results} about here
------
}
\hspace{1pt}

\input{tables/simulations}

Nevertheless, although controlling for $Z_1$ is sufficient to obtain a consistent estimate for $X$, its marginal effect will itself not correspond to any causal effect of $Z_1$ on $Y$. That is because $Z_1$ is correlated with $Z_2$ and will thus partially pick up an effect of $Z_2$ on $Y$ too \parencite{Cinelli2020}. To illustrate this phenomenon quantitatively, we parameterize the causal graph in Figure \ref{fig:1a} in the following way:
\begin{equation}
\label{eq1}
    \begin{aligned}
        z_1 &\leftarrow u + \varepsilon_1, \\
        z_2 &\leftarrow u + \varepsilon_2, \\
        x &\leftarrow z_1 + \varepsilon_3, \\
        y &\leftarrow x + z_2 + \varepsilon_4,
    \end{aligned}
\end{equation}
with $n=10,000$, and $U$, $\varepsilon_i$ being standard normal. True effect sizes are set equal to one. Note that $U$ is assumed to be unobserved and appears in the functions assigning values to $Z_1$ and $Z_2$. This creates an error correlation between the two variables. We then run a regression of $Y$ on $X$ and $Z_1$, which gives a consistent coefficient estimate for $X$ ($\hat{\beta}_X$ = 1.017), while the effect of $Z_1$  ($\hat{\beta}_{Z_1}$ = 0.499) turns out to be biased. By contrast, if we also include $Z_2$ in the regression, the coefficient of $Z_1$ drops to zero ($\hat{\beta}_{Z_1}$ = -0.019), which corresponds to its actual causal effect on $Y$ in this example (since $Z_1$ does not appear in line 4 of eq. \ref{eq1}). Detailed simulation results together with associated standard errors (bootstrapped with 1,000 replications) are reported in Table \ref{tab:sim_results}.

Figures \ref{fig:1b} and \ref{fig:1c} highlight under which conditions effect estimates of control variables can be interpreted causally. In Figure \ref{fig:1b} there are two backdoor paths: $X \leftarrow Z_1 \rightarrow Y$ and $X \leftarrow Z_1 \dashleftarrow\dashrightarrow Y$. Both paths can be intercepted by $Z_1$, which is thus a valid control variable. Data simulated according to the following system:
\begin{equation}
\label{eq2}
    \begin{aligned}
        z_1 &\leftarrow u + \varepsilon_1, \\
        x &\leftarrow z_1 + \varepsilon_2, \\
        y &\leftarrow x + z_1 + u + \varepsilon_3,
    \end{aligned}
\end{equation}
with coefficients again set equal to one, confirm that the causal effect of $X$ can be consistently estimated in a regression of $Y$ on $X$ and $Z_1$ ($\hat{\beta}_X$ = 0.993). However, once again the coefficient estimate for $Z_1$ ($\hat{\beta}_{Z_1}$ = 1.503) is biased (Table 1, col. 4). Although $Z_1$ is a valid control variable in eq.\ \ref{eq2}, it is nonetheless endogenous \parencite{Froelich2008}. Note that the unobserved variable $U$ enters the combined error term $\nu = u + \varepsilon_3$ in line 3 of eq.\ \ref{eq2}. At the same time, $U$ is an argument of the function assigning values to $Z_1$ (line 1 of eq.\ \ref{eq2}), which lets $Z_1$ become correlated with the error term.\footnote{For ease of exposition, all coefficients in our simulations are set equal to one. However, it is easy to construct examples where the sign of the control variable $Z_1$ changes; e.g., by setting the coefficient of $U$ equal to $-3$ in line 3 of eq.\ \ref{eq2}, which results in $\hat{\beta}_{Z_1} = -0.518$ (std.\ err.\ = 0.029).}

This is different in Figure \ref{fig:1c}. Here the two backdoor paths are $X \leftarrow Z_1 \rightarrow Y$ and $X \dashleftarrow\dashrightarrow Z_1 \rightarrow Y$. When we simulate data according to:
\begin{equation}
\label{eq3}
    \begin{aligned}
        z_1 &\leftarrow u + \varepsilon_1, \\
        x &\leftarrow z_1 + u + \varepsilon_2, \\
        y &\leftarrow x + z_1 + \varepsilon_3,
    \end{aligned}
\end{equation}
we now find that a regression of $Y$ on $X$ and $Z_1$ provides a consistent estimate for both $X$ as well as for $Z_1$ ($\hat{\beta}_X$ = 1.001; $\hat{\beta}_{Z_1}$ = 1.004; Table 1, col. 5). In this situation, the regression coefficient for the control variable $Z_1$ has a causal interpretation. This is because, unlike in the previous situations, we are able to account for all influence factors of $Y$, apart from the exogenous error term $\varepsilon_3$. In particular, there is no unobserved variable $U$ jointly affecting the outcome $Y$ and (at least one of) the regressors $(X, Z_1)$ anymore. 

Finally, Figure \ref{fig:1d} depicts a more complex setting, with several admissible sets of controls, each sufficient to identify the causal effect of $X$ on $Y$ \parencite{Textor2011}. One possibility in this situation is to simply control for $Z_1$, which is the only direct influence factor of $X$, and thus blocks all paths entering $X$ through the backdoor. To witness, we simulate data from the system:
\begin{equation}
\label{eq4}
    \begin{aligned}
        z_1 &\leftarrow u_1 + \varepsilon_1, \\
        z_2 &\leftarrow z_1 + u_1 + u_2 + \varepsilon_2, \\
        z_3 &\leftarrow z_2 + \varepsilon_3, \\
        z_4 &\leftarrow z_2 + \varepsilon_4, \\
    \end{aligned}
    \qquad
    \begin{aligned}
        z_5 &\leftarrow z_2 + u_2 + \varepsilon_5, \\
        x &\leftarrow z_1 + \varepsilon_6, \\
        y &\leftarrow x + z_3 + z_4 + z_5  + \varepsilon_7,
    \end{aligned}
\end{equation}
and regress $Y$ on $X$ and $Z_1$, which gives a consistent estimate for the effect of $X$ ($\hat{\beta}_X$ = 0.991; Table 1, col. 6). Similarly, controlling for the direct influence factors of $Y$ ($Z_3$, $Z_4$, and $Z_5$) also blocks all backdoor paths and leads to a consistent effect estimate for $X$ ($\hat{\beta}_X$ = 1.006; Table 1, col. 7). A third alternative is to control for the entire set of covariates ($Z_1$, $Z_2$, $Z_3$, $Z_4$, and $Z_5$) which also leads to a consistent estimate for $X$ ($\hat{\beta}_X$ = 1.003; Table 1, col. 8), although this would be the most data-intensive identification strategy leading to slightly less precise estimates compared to the previous specification, due to fewer degrees of freedom. This example illustrates that the minimally sufficient set of controls (here: $Z_1$) for identifying the causal effect of $X$ is often much smaller than the total number of confounding variables in a model. At the same time, the estimated marginal effects for the control variables only have a causal interpretation if all the direct influence factors of $Y$ (here: $Z_3$, $Z_4$, and $Z_5$) are accounted for in the regression. As we argued above, this is unlikely to be the case, since in many real-world settings the number of causal factors determining $Y$ might be prohibitively large.\footnote{Alternatively, if omitted factors are unrelated to all other regressors, a causal interpretation of control variable coefficients would be possible too. This seems likewise implausible in many applications though.}

%% file: figures/figure1.tex
\begin{figure}[thp]
    \centering
        \begin{subfigure}[t]{0.47\textwidth}
        \centering
            \begin{tikzpicture}[>={Latex[length=2mm,width=1.5mm]}, font=\footnotesize]
            \node[fill,circle,inner sep=0pt,minimum size=5pt,label={below:{X}}] (X) at (0,0) {};
    		\node[fill,circle,inner sep=0pt,minimum size=5pt,label={below:{Y}}] (Y) at (5,0) {};
    		\node[fill,circle,inner sep=0pt,minimum size=5pt,label={left:{Z$_1$}}] (Z1) at (0.8,1.7) {};
    		\node[fill,circle,inner sep=0pt,minimum size=5pt,label={right:{Z$_2$}}] (Z2) at (4.2,1.7) {};
    		\draw[->,shorten >= 1pt] (X)--(Y);
    		\draw[->,shorten >= 1pt] (Z1)--(X);
    		\draw[->,shorten >= 1pt] (Z2)--(Y);
    		\draw[<->,dashed,shorten >= 1pt] (Z1) to[bend left=45] (Z2);
                \end{tikzpicture}
                \caption{}
                \label{fig:1a}
        \end{subfigure}
        \begin{subfigure}[t]{0.47\textwidth}
            \centering
            \begin{tikzpicture}[>={Latex[length=2mm,width=1.5mm]}, font=\footnotesize]
            \node[fill,circle,inner sep=0pt,minimum size=5pt,label={below:{X}}] (X) at (0,0) {};
    		\node[fill,circle,inner sep=0pt,minimum size=5pt,label={below:{Y}}] (Y) at (5,0) {};
    		\node[fill,circle,inner sep=0pt,minimum size=5pt,label={above:{Z$_1$}}] (Z1) at (2.5,2) {};
    		\draw[->,shorten >= 1pt] (X)--(Y);
    		\draw[->,shorten >= 1pt] (Z1)--(X);
            \draw[->,shorten >= 1pt] (Z1)--(Y);
    		\draw[<->,dashed,shorten >= 1pt] (Z1) to[bend left=60] (Y);
                \end{tikzpicture}
                \caption{}
                \label{fig:1b}
        \end{subfigure}
        \begin{subfigure}[t]{0.47\textwidth}
            \centering
            \begin{tikzpicture}[>={Latex[length=2mm,width=1.5mm]}, font=\footnotesize]
            \node[fill,circle,inner sep=0pt,minimum size=5pt,label={below:{X}}] (X) at (0,0) {};
    		\node[fill,circle,inner sep=0pt,minimum size=5pt,label={below:{Y}}] (Y) at (5,0) {};
    		\node[fill,circle,inner sep=0pt,minimum size=5pt,label={above:{Z$_1$}}] (Z1) at (2.5,2) {};
    		\draw[->,shorten >= 1pt] (X)--(Y);
    		\draw[->,shorten >= 1pt] (Z1)--(X);
            \draw[->,shorten >= 1pt] (Z1)--(Y);
    		\draw[<->,dashed,shorten >= 1pt] (X) to[bend left=60] (Z1);
                \end{tikzpicture}
                \caption{}
                \label{fig:1c}
        \end{subfigure}
        \begin{subfigure}[t]{0.47\textwidth}
        \centering
            \begin{tikzpicture}[>={Latex[length=2mm,width=1.5mm]}, font=\footnotesize]
            \node[fill,circle,inner sep=0pt,minimum size=5pt,label={below:{X}}] (X) at (0,0) {};
    		\node[fill,circle,inner sep=0pt,minimum size=5pt,label={below:{Y}}] (Y) at (6,0) {};
    		\node[fill,circle,inner sep=0pt,minimum size=5pt,label={left:{Z$_1$}}] (Z1) at (0.8,1.7) {};
    		\node[fill,circle,inner sep=0pt,minimum size=5pt,label={[yshift=0.1cm]above:{Z$_2$}}] (Z2) at (3,3) {};
    		\node[fill,circle,inner sep=0pt,minimum size=5pt,label={left:{Z$_3$}}] (Z3) at (3,1) {};
    		\node[fill,circle,inner sep=0pt,minimum size=5pt,label={left:{Z$_4$}}] (Z4) at (4.5,1.5) {};
    		\node[fill,circle,inner sep=0pt,minimum size=5pt,label={right:{Z$_5$}}] (Z5) at (5.7,1.9) {};
    		\draw[->,shorten >= 1pt] (X)--(Y);
    		\draw[->,shorten >= 1pt] (Z1)--(X);
    		\draw[->,shorten >= 1pt] (Z1)--(Z2);
    		\draw[->,shorten >= 1pt] (Z2)--(Z3);
    		\draw[->,shorten >= 1pt] (Z2)--(Z4);
    		\draw[->,shorten >= 1pt] (Z2)--(Z5);
    		\draw[->,shorten >= 1pt] (Z3)--(Y);
    		\draw[->,shorten >= 1pt] (Z4)--(Y);
    		\draw[->,shorten >= 1pt] (Z5)--(Y);
    		\draw[<->,dashed,shorten >= 1pt] (Z1) to[bend left=60] (Z2);
    		\draw[<->,dashed,shorten >= 1pt] (Z2) to[bend left=60] (Z5);
                \end{tikzpicture}
                \caption{}
                \label{fig:1d}
        \end{subfigure}
    \caption{Examples of causal diagrams with valid control variable $Z_1$}
    \label{fig:1}
\end{figure}

%% file: tables/simulations.tex
\begin{table}
\caption{OLS regressions with varying adjustment sets} 
\label{tab:sim_results}
\begin{tabular}{p{1cm}cccccccc}
\toprule
\addlinespace
   & \multicolumn{3}{c}{Figure \ref{fig:1a}} & Figure \ref{fig:1b} & Figure \ref{fig:1c} & \multicolumn{3}{c}{Figure \ref{fig:1d}}\\ \cmidrule(rl){2-4} \cmidrule(rl){5-5} \cmidrule(rl){6-6} \cmidrule(rl){7-9}
   & (1) & (2) & (3) & (4) & (5) & (6) & (7) & (8) \\ 

   \addlinespace
   \midrule 
   \addlinespace
   \multicolumn{9}{l}{\underline{Treatment Variable:}}\\
    \addlinespace
    \addlinespace
                          
   $X$ & 1.017 & 1.004 & 1.015 & 0.993 & 1.001 & 0.991 & 1.006 & 1.003 \\
         & (0.015) & (0.006) & (0.010) & (0.012) & (0.008) & (0.057) & (0.007) & (0.010) \\

    \addlinespace
    \midrule
    \addlinespace
    \multicolumn{9}{l}{\underline{Control Variables \textbf{(not to be interpreted causally)}:}}\\
    \addlinespace
    \addlinespace
   
   $Z_1$ & 0.499 & & -0.019 & 1.503 & 1.004 & 4.565 & & 0.004 \\
         & (0.018) & & (0.013) & (0.014) & (0.014) & (0.069) & & (0.016) \\
   
   $Z_2$ & & 0.993 & 0.997 & & & & & 0.009 \\
         & & (0.008) & (0.008) & & & & & (0.019) \\
   
   $Z_3$ & & & & & & & 0.994 & 0.991 \\
         & & & & & & & (0.008) & (0.010) \\
   
   $Z_4$ & & & & & & & 0.991 & 0.988 \\
         & & & & & & & (0.008) & (0.010) \\
   
   $Z_5$ & & & & & & & 1.011 & 1.009 \\
         & & & & & & & (0.006) & (0.008) \\ \addlinespace
   \bottomrule
   \multicolumn{9}{p{.9\textwidth}}{\footnotesize \emph{Note:} Simulation results (as discussed in the main text) using different backdoor-admissible adjustment sets for the causal models depicted in Figure \ref{fig:1}a--d. Bootstrapped standard errors (with $1,000$ replications and $n = 10,000$) in parentheses. True effect sizes for all variables equal to one.}
\end{tabular}
\end{table}

%% file: 4_example.tex
\section{Examples}

In the following section we present practical applications that illustrate our previous theoretical points. To start as simply as possible, we were looking for examples that employ standard regression models instead of more advanced empirical techniques. One challenge in this regard is that papers with simple OLS regressions often refrain from making causal claims and instead resort to alternative formulations to describe effect sizes, such as "association", "pattern", or "link" \parencite{Hernan2018}. However, a recent paper by \Textcite{Hoffman2023} provides a fitting example. The authors estimate the causal effect of longer travel time on the probability of a default judgement being made in eviction cases as a result of defendants not showing up or being late to court. Using OLS in a sample of more than 200,000 eviction proceedings in the city of Philadelphia between 2005 and 2021, they find that an increase of one hour in estimated travel time raises the likelihood of a default judgement by 3.8 to 8.6 percent. This effect is meaningful because defaults are difficult to reopen and tenants who fail to show up in court cannot benefit from ``Civil Gideon" protections offered in major urban areas in the U.S.

In their models, Hoffman and Strezhnev control for neighborhood characteristics such as census tract income levels, as well as race and ethnicity. Interestingly, for our discussion, they also control for building characteristics and find a positive and statistically significant effect size of multi-unit apartment buildings (compared to row houses or single family dwelling) on the probability of default judgements. However, as the authors discuss in their paper, this relationship is unlikely to have a causal interpretation, since building characteristics might be correlated with other influence factors such as unfavorable terms in residential leases or the geographical distribution of dwellings within the city. We now turn to an example closer to the organizational research context.

\subsection{Early research exposure and career choices}

\Textcite{Azoulay2021} investigate the effect of early career exposure to frontier research on the career trajectory of potential innovators. Their specific empirical setting is the Associate Training Program (ATP) of the National Institutes of Health (NIH) in the U.S. The ATP was started in 1953 as a training program for recent MD graduates. Participants were sent to the NIH intramural campus in Bethesda, Maryland, to receive a two to three years research training under the supervision of NIH investigators. Since the NIH was originally established within the Marine Hospital Service, participation in the program fulfilled a draftee’s military service obligation. Therefore, applying to the ATP became particularly popular among young physicians during the Vietnam War period (1965--1975).

After a first screening round, applicants were invited to an interview on NIH campus to determine who would eventually be selected to participate in the program. Selection criteria were related to applicants' prior research activities (which Azoulay et al.\ measure by their number of pre-ATP publications), their academic achievements (proxied by whether they were elected to the $\mathrm{A} \Omega \mathrm{A}$ Honor Medical Society), experience (i.e., whether they held a Ph.D.\ at the time of application and the number of internships they had completed), and the reputation of the institutions where applicants received their training (measured by NIH grants for applicants' medical school and internship hospital). Importantly, Azoulay et al. argue that although the pool of applicants to the ATP was indeed a highly selected group, selection at the (second) interview stage was based entirely on these observable characteristics. Applicants were early in their career and rather homogeneous in their characteristics. It was therefore hard to select them based on their future research potential beyond a few observable markers. This feature of the particular institutional setting allows Azoulay et al. to employ a selection-on-observables design. Based on that they estimate that ATP participants were twice as likely to pursue a research-focused career later on compared to unsuccessful applicants. As a result, trainees accumulated more publications, citations and grant funding over their life-cycle. Furthermore, they were significantly more likely to receive prestigious career awards, including the Nobel Prize, and to become elected members of the National Academy of Sciences.

\hspace{1pt}
\centerline{
------
Insert Figure \ref{fig:2} about here
------
}
\hspace{1pt}

\input{figures/figure2}

Figure \ref{fig:2} synthesizes the assumptions leading to the empirical strategy in Azoulay et al. in form of a causal diagram. Controlling for applicants' prior research activities, academic achievements, experience and school reputation (the authors incorporate several covariates for each of these dimensions, including medical school and internship hospital fixed effects) is a valid backdoor adjustment set for estimating the causal effect of ATP participation on the choice of pursuing a research career in this graph.\footnote{For simplicity, we focus on career choice as the primary outcome and disregard several other dimensions of scientific career success that Azoulay et al. investigate.} The analysis depends crucially on the assumption that the unobserved (latent) variable \emph{research potential} does not directly affect program participation (ATP Participation \ $\not\!\!\!\dashleftarrow$ Research Potential); i.e., interviewers are not able to select applicants based on private information.

Azoulay et al. employ an inverse probability weighting estimator \parencite{Austin2015}. As such, covariates are only used to estimate the propensity score of receiving treatment and do not appear in an outcome regression. However, in this setting it would also not be advisable to interpret the effect of control variables such as prior research activities on career choice. The latent node research potential jointly affects an applicant's prior research activities as well as future career choices. Thus, while prior research activity is a valid control for the effect of ATP participation, it is also endogenous, similar to the situation in Figure \ref{fig:1b}. Consequently, even if we were to find a positive correlation between prior research activities and pursuing a research career (which is not reported in Azoulay et al.), it would be premature to conclude that, e.g., early publication success during medical school is a significant driver of subsequent career choices, since both of these variables are likely confounded by an applicant's overall ability. The research design only allows to draw policy conclusions for the treatment variable ATP participation. Researchers should therefore be careful not to overinterpret their empirical results, even if that promises to offer interesting additional perspectives on a given research topic.

\subsection{Analyst coverage and innovation}

In applied empirical research, it is not uncommon for estimated treatment effects to change significantly when more advanced identification strategies are employed compared to standard OLS regression. For example, \Textcite{Hopp2020} find that CEO appearance is no longer related to company performance once firm fixed effects are incorporated in the analysis. Using a discrete choice experiment, \Textcite{Mas2017} show that on average workers value flexible work arrangements much less than a simple compensating wage differentials regression would indicate. Furthermore, because of simultaneity bias, policing levels and crime rates are often positively related, while \Textcite{Mello2018} demonstrates a negative causal effect of policing on crime, exploiting a natural experiment in a difference-and-difference design.

If subsequent research uses the same variables as controls, however, it becomes immediately clear that their estimated regression coefficients should not be interpreted in a causal way. One case in point comes from the literature on analyst coverage and innovation. \Textcite{He2013} find a negative relationship between analyst coverage and patenting in a study of U.S.\ public firms from 1993 to 2005, using difference-and-differences and an instrumental variable approach. It has been theorized that this result arises because external stock market analysts following a firm often exert excessive pressure on executives, which can worsen managerial myopia and impede investment in long-term innovation projects. For these reasons, analyst coverage is used as a control variable in other studies of R\&D activities in publicly listed firms. However, since it is not the main variable of interest in these studies, often less stringent identification strategies are employed, possibly leading to unexpected results. For example, \Textcite{Chen2016} and \Textcite{Huang2022} consistently find positive effects of analyst coverage in regressions with the natural logarithm of patents as the dependent variable, which seemingly contradicts \Textcite{He2013}.

Analyst coverage can be a valid control variable even though it is endogenous (if the underlying causal structure, e.g., corresponds to Figure \ref{fig:1b}). Nevertheless, interpreting positive regression coefficients as evidence against \Textcite{He2013} would be a mistake. To use the analogy of Bayesian updating, control variable estimates from studies such as \Textcite{Chen2016} and \Textcite{Huang2022} should not alter the posterior probability of analyst coverage having a negative effect on innovation. Consequently, in another study by \Textcite{Chen2021}, in which analyst coverage is also included as a control variable in a patent count regression, the regression coefficients of the controls are not reported. This is in accordance with the recommendations that we will discuss in the following.

%% file: figures/figure2.tex
\begin{figure}[thp]
\centering
    \begin{tikzpicture}[>={Latex[length=2mm,width=1.5mm]}, font=\footnotesize]
        \node[fill,circle,inner sep=0pt,minimum size=5pt,label={below:{ATP Participation}}] (X) at (0,0) {};
    	\node[fill,circle,inner sep=0pt,minimum size=5pt,label={below:{Career Choice}}] (Y) at (9.5,0) {};
    	\node[fill,circle,inner sep=0pt,minimum size=5pt,label={[align=center]Prior Research\\Activities}] (Z1) at (1,4) {};
        \node[fill,circle,inner sep=0pt,minimum size=5pt,label={[align=center]Academic\\Achievements}] (Z2) at (3.5,4) {};
    	\node[fill,circle,inner sep=0pt,minimum size=5pt,label={[align=center]Experience}] (Z3) at (6,4) {};
        \node[fill,circle,inner sep=0pt,minimum size=5pt,label={[align=center]School\\Reputation}] (Z4) at (8.5,4) {};
        \node[draw,fill=white,circle,inner sep=1.5pt,outer sep=0,minimum size=5pt,label=right:{\parbox[c]{3.0cm}{Research\\Potential}}] (U) at (10.5,1.5) {};
    	\draw[->,shorten >= 1pt] (X)--(Y);
    	\draw[->,shorten >= 1pt] (Z1)--(X);
    	\draw[->,shorten >= 1pt] (Z1)--(Y);
        \draw[->,shorten >= 1pt] (Z2)--(X);
    	\draw[->,shorten >= 1pt] (Z2)--(Y);
        \draw[->,shorten >= 1pt] (Z3)--(X);
    	\draw[->,shorten >= 1pt] (Z3)--(Y);
        \draw[->,shorten >= 1pt] (Z4)--(X);
    	\draw[->,shorten >= 1pt] (Z4)--(Y);
    	\draw[->,dashed,shorten >= 1pt] (U)--(Z1);
        \draw[->,dashed,shorten >= 1pt] (U)--(Z2);
        \draw[->,dashed,shorten >= 1pt] (U)--(Z3);
        \draw[->,dashed,shorten >= 1pt] (U)--(Z4);
        \draw[->,dashed,shorten >= 1pt] (U)--(Y);
    \end{tikzpicture}
\caption{Causal diagram visualizing the empirical strategy in \Textcite{Azoulay2021}}
\label{fig:2}
\end{figure}

%% file: 5_discussion.tex
\section{Discussion and Recommendations}

Beyond pure prediction tasks, the purpose of regression analysis in organizational research is typically to build and test theories that explain the causal mechanisms underlying a studied phenomenon \parencite{Sutton1995}.  In this paper, we argued that attaching substantive meaning to the marginal effects of biased control variables is problematic, however, as researchers could develop false intuitions or draw erroneous managerial and policy conclusions. Therefore, we think it is advisable to not discuss the results obtained for control variables in quantitative papers, unless the researchers can be sure that they have accounted for all relevant influence factors of the outcome in a regression (\emph{all-causes regression}). Since in many practical settings this is unlikely to be the case, we recommend to treat controls as \emph{nuisance parameters}, which are included in the analysis for identification purposes (and discussed as such) but their effects are not interpreted \parencite{Liang1995, Meehl1971}. This corresponds to the way control variables are treated by non-parametric matching estimators \parencite{Heckman1998} and modern machine learning techniques for high-dimensional settings \parencite{Chernozhukov2018,Huenermund2023b}. These methods do not report estimation results related to controls, either because there would be simply too many covariates in the analysis (which is the primary use-case for machine learning) or marginal effects of control variables are not returned by the estimation protocol (as in the matching case).

Our recommendations thereby depart from prior literature insofar as control variable should not be promoted to have equal status with the other variables in the study \parencite[p.\ 297]{Spector2011}. Research designs based on control variables are employed to estimate the causal effect of a treatment variable on an outcome. As such, the treatment variable cannot be endogenous, otherwise estimates would be biased and other, more suitable research designs (such as instrumental variables, regression discontinuity designs, etc.) should be applied. By contrast, control variables can be endogenous \parencite{Froelich2008} and, as we argued in the preceding theoretical discussion, will likely be so in practice.\footnote{Similarly, the widespread notion that one endogenous variable in the regression would bias all other coefficients is also incomplete, as our simulation of the causal model in Figure \ref{fig:1b} (Table \ref{tab:sim_results}, col.\ 4) demonstrates.} Controls should be chosen to close all backdoor paths between a treatment and outcome, based on a theoretical model of the context under study \parencite{Bono2011}. As we have demonstrated previously, it is thereby not necessary to include all causal influence factors of the outcome variable in a regression. Our example \parencite{Azoulay2021} illustrates that in many cases it might actually be easier to control the treatment assignment mechanism instead, if institutional knowledge is richer about what determines treatment take-up compared to the potentially long list of variables that affect the outcome. Moreover, in many situations, researchers have the choice between different valid adjustment sets (see Figure \ref{fig:1d}), which highlights their auxiliary nature for the analysis.

\hspace{1pt}
\centerline{
------
Insert Table \ref{tab:proposed} about here
------
}
\hspace{1pt}

\input{tables/proposed_table}

Since accounting for all influence factors of the outcome might be unrealistic in many contexts and control variables are therefore likely to be endogenous, interpreting their effect sizes in light of theory is potentially dangerous. Authors could infer wrong conclusions for managerial advice and subsequent studies might be inclined to build theory based on biased empirical results. To avoid this, we therefore recommend to refrain from interpreting control variables in published papers. Moreover, predicting the sign of control variable estimates ex-ante \parencite{Atinc2012} is difficult if endogenous control variables can pick up the effect of a multitude of other influence factors. Therefore, formulations such as ``estimates of control variables have expected signs" should be avoided. As a ``nudge" to stir the research community away from overinterpreting control variables, we find it appropriate that authors omit their coefficients entirely from regression tables or relegate them to an appendix. 

Table \ref{tab:proposed} (which is an adapted version of Table \ref{tab:sim_results}) illustrates such a regression table format, in which check marks are included to indicate which variables were controlled for. This corresponds to the way how estimation results are presented, e.g., in papers using nearest neighbor or propensity score matching. We acknowledge that omitting control variable coefficients in regression tables constitutes a trade-off with respect to transparency. However, we believe that this suggestion is justified by the lower risk of drawing incorrect theoretical inferences from empirical studies. When authors have important reasons for reporting regression coefficients of control variables, the format of Table \ref{tab:sim_results} constitutes a viable compromise, in our view. Compared to the standard format of regression table,  Table \ref{tab:sim_results} clearly separates controls from the treatment variable and includes a note that the control variable estimates should not be interpreted causally. 

We emphasize that we agree with \Textcite{Becker2005} in that control variables should be carefully discussed and authors need to justify their validity based on prior theory. However, their estimated coefficients are less relevant. It suffices to discuss the rationale for selecting specific control variables in the empirical section and to clearly indicate their inclusion in the table notes. Since the proper justification of a regression design can only come from theory, we caution against deciding about the inclusion of control variables based on their incremental contribution to the $R^2$ of the model \parencite{Carlson2012}. This is the celebrated \emph{"no causes in, no causes out"} principle \parencite{Cartwright1989}, which states that the validity of causal inferences must ultimately be supported by theoretical considerations external to the data. For example, \emph{bad controls} (as discussed in \Textcite{Cinelli2022}) often have a lot of predictive power but nonetheless lead to invalid causal inferences. Therefore, we also do not see a reason why authors should report models with and without control variables and compare the share of explained variance between them (as suggested, e.g., by \Textcite{Atinc2012} and \Textcite{becker2016}).

Our recommendations are in line with \Textcite{Westreich2013} who discuss a similar problem with respect to the interpretation of potentially endogenous controls in epidemiology. Because epidemiological studies usually present the results of multivariate regression analyses right after a table with descriptive statistics of the data, they coined the term \emph{table 2 fallacy}. \Textcite{Keele2020} discuss related examples from the field of political science. They emphasize that for estimates of control variables to be given a causal interpretation, their effects need to be themselves causally identified. Since this is only plausible if there are no omitted variables (or the controls are unrelated to the omitted variables), we recommend researchers to focus attention on one causal factor (or a small set) at a time, for which backdoor paths can realistically be enumerated, and treat control variables as nuisance parameters instead.
 
Finally, we caution against including estimates of potentially biased controls in meta-analyses \parencite{Anguinis2011}. Such studies pool the effects of a focal variable on an outcome across several papers. According to \Textcite{Becker2005}, systematic reporting of control variables facilitates cumulative science and knowledge aggregation by significantly increasing the pool of studies from which effect sizes for meta-analyses can be drawn from:

\begin{displayquote}
``Nonreporting of control variable findings hinder any meta-analyses that would have otherwise included the controls. For instance, in a study of the relationship between employee commitment and organizational citizenship behavior, a researcher might control for extraversion and agreeableness but not report the findings for the controls. As a result, later meta-analyses cannot include these findings in the assessment of connections between personality and organizational citizenship behavior." \parencite[p.\ 285]{Becker2005}
\end{displayquote}
This recommendation refers to meta-analyses of partial correlations and marginal effects (``meta-regression"), which are increasingly common in organizational research and economics \parencite{Stanley2012}. Compared to zero-order correlations, they have the advantage of being able to filter out other potential confounding influence factors in settings when randomized control trials are not feasible.\footnote{See \Textcite{DeSimone2021} for methodological guidelines on how to conduct meta-analytic studies in organizational research.} However, the quoted passage fails to mention that control variables (here: extraversion and agreeableness) are unlikely to have a causal interpretation themselves and therefore add little to the evidence base regarding a certain effect size. As their coefficients may represent a combination of several different causal mechanisms jointly operating on the outcome (here: citizenship behavior), they do not provide accurate information about a theoretically  meaningful quantity. Moreover, coefficients can vary substantially depending on which admissible adjustment sets are used (e.g., compare columns\ 6--8 in Table \ref{tab:sim_results}). Consequently, meta-analyses should be restricted to the main treatment variable(s), for which a plausible identification argument can be established, which highlights once again the unequal status of treatment and control variables in regression analysis.

To conclude, there is no reason to be worried if the estimated coefficients of control variables do not have expected signs, since they are likely to be biased anyway in practical applications. Instead, researchers should rather focus on interpreting the marginal effects of the main variables of interest in their manuscripts. The estimation results obtained for controls, by contrast, have little substantive meaning and can therefore safely be omitted---or relegated to an appendix. This approach will not only prevent researchers from drawing wrong causal conclusions based on endogenous controls, but will furthermore allow to streamline the discussion sections of quantitative research papers and save on valuable manuscript space.

%% file: tables/proposed_table.tex
\begin{table}
\caption{Regression table without control variable coefficients being reported} 
\label{tab:proposed}
\begin{tabular}{p{1cm}cccccccc}
\toprule
\addlinespace
   & \multicolumn{3}{c}{Figure \ref{fig:1a}} & Figure \ref{fig:1b} & Figure \ref{fig:1c} & \multicolumn{3}{c}{Figure \ref{fig:1d}}\\ \cmidrule(rl){2-4} \cmidrule(rl){5-5} \cmidrule(rl){6-6} \cmidrule(rl){7-9}
   & (1) & (2) & (3) & (4) & (5) & (6) & (7) & (8) \\ 

   \addlinespace
   \midrule 
    \addlinespace
    \addlinespace
                          
   $X$ & 1.017 & 1.004 & 1.015 & 0.993 & 1.001 & 0.991 & 1.006 & 1.003 \\
         & (0.015) & (0.006) & (0.010) & (0.012) & (0.008) & (0.057) & (0.007) & (0.010) \\

    \addlinespace
    \midrule
    \addlinespace
   $Z_1$ & $\checkmark$ & & $\checkmark$ & $\checkmark$ & $\checkmark$ & $\checkmark$ & & $\checkmark$ \\
   $Z_2$ & & $\checkmark$ & $\checkmark$ & & & & & $\checkmark$ \\
   $Z_3$ & & & & & & & $\checkmark$ & $\checkmark$ \\
   $Z_4$ & & & & & & & $\checkmark$ & $\checkmark$ \\
   $Z_5$ & & & & & & & $\checkmark$ & $\checkmark$ \\
   \bottomrule
   \multicolumn{9}{p{.9\textwidth}}{\footnotesize \emph{Note:} Bootstrapped standard errors in parentheses.}
\end{tabular}
\end{table}

%% file: tables/appendix_smj.tex

\begin{longtable}[!]{ccccp{9cm}}
\caption{Strategic Management Journal} \\
\toprule
\textbf{Volume} & \textbf{Issue} & \textbf{Total} & \textbf{Count} & \textbf{Articles} \\* \midrule
\endhead
\bottomrule
\endfoot
\endlastfoot
41 & 13 & 6 & 2 & Polidoro (2020); Wiersema, Ahn \& Zhang (2020);                                 \\
41 & 12 & 4 & 2 & Kiss, Libaers, Barr, Wang \& Zachary (2020); Zheng \& Wang (2020);              \\
41 & 11 & 5 & 2 & Argyres, Rios \& Silverman (2020); Zhou \& Park (2020);                         \\
41 & 10 & 4 & 3 & Rawley \& Seamans  (2020); Uribe (2020); Zhu, hu \& Shen (2020);                 \\
41 & 9  & 6 & 3 & Lee (2020); Chahine \& Zhang (2020); Blagoeva, Kuvasan \& Jansen (2020);         \\
41 & 8  & 4 & 2 & Dutt \& Mitchel (2020); Moeen \& Mitchel (2020);                                \\
41 & 7  & 7 & 3 & Belderbos, Tong \& Wu (2020); Hoon Oh, Shapiro, Ho \& Shin (2020); Skiti (2020); \\
41 & 7 & 7 & 2 & Belderbos, Tong \& Wu (2020); Oh, Shapiro, Ho \& Shin (2020); \\
41 & 6 & 6 & 2 & Chattopadhyay \& Bercovitz (2020); Smulowitz, Rousseau \& Bromiley (2020); \\
41 & 5 & 4 & 2 & Sakakibara \& Balasubramanian (2020); Rocha \& van Praag, (2020); \\
41 & 4 & 7 & 2 & Aggarwal (2020); Bonet, Capelli \& Homari (2020); \\
41 & 3 & 4 & 2 & Arikan, Arikan \& Shenkar (2020); Agarwal, Braguinsky \&   Ohyama (2020); \\
41 & 2 & 5 & 2 & Ryu, Reuer \& Brush (2020); Jia, Gao \& Julian (2020); \\
41 & 1 & 5 & 0 & - \\
40 & 13 & 7 & 1 & Hsu, Kov\'{a}cs \& Ko\c{c}ak (2019); \\
40 & 12 & 6 & 4 & Kim (2019); Petrenko, Aime, Recendes \& Chandler (2019);   Guldiken, Mallon, Fainshmidt, Judge \& Clark (2019); Shi, Conelly, Mackey   \& Gupta (2019); \\
40 & 11 & 5 & 2 & Woo, Canella \& Mesquita (2019); Zweiger, Stettler, Baldauf \& Zamudio (2019); \\
40 & 10 & 6 & 3 & Ridge, Imgram, Abdurakhmonov \& Hasija (2019);   G\'{o}mez‐-Sol\'{o}rzano, Tortoriello \& Soda (2019); Kavusan \& Frankort (2019); \\
40 & 9 & 5 & 0 & - \\
40 & 8 & 6 & 1 & Barlow, Verhaal \& Angus (2019); \\
40 & 7 & 5 & 2 & Corsino, Mariani \& Torrisi (2019); Andrus, Withers,   Courtright \& Boivie (2019); \\
40 & 6 & 5 & 2 & Hiatt \& Carlos (2019); Piazzai \& Wijnberg (2019); \\
40 & 5 & 4 & 2 & Hill, Recendes \& Ridge (2019); Yu, Minniti \& Nason (2019); \\
40 & 4 & 5 & 3 & Paik, Kang \& Seamans (2019); Bruce, de Figueiredo \&   Silverman (2019); Zheng, Ni \& Crilly (2019); \\
40 & 3 & 3 & 2 & Chatterji, Delecourt, Hasan \& Koning (2019); Bigelow,   Nickerson \& Park (2019); \\
40 & 2 & 5 & 3 & Criscuolo, Alexy, Sharapov \& Salter (2019); Ren, Hu \&   Cui (2019); Boone, Lokshin, Guenter \& Belderbos (2019); \\
40 & 1 & 7 & 4 & Haans (2019) (Appendix); Chatterji, Cunningham \& Joseph (2019); Westphal \& Zhu (2019); Belderbos, Tong \& Wu (2019); \\
39 & 13 & 5 & 1 & Garg \& Zhao (2018); \\
39 & 12 & 5 & 4 & Cui, Yang \& Vertinsky (2018) (Appendix); Ranganathan, Ghosh   \& Rosenkopf (2018); Arslan (2018); Asgari, Tandon, Singh \& Mitchell,   2018; \\
39 & 11 & 8 & 5 & Feldman, Gartenberg \& Wulf (2018); Claussen, Essling \&   Peukert (2018); Burbano, Mamer \& Snyder (2018); Koch-Bayram \& Wernicke,   2018; Mata \& Alves (2018); \\
39 & 10 & 8 & 4 & Eberhardt \& Eesley (2018); Hornstein \& Zhao (2018); Kang   \& Zaheer (2018); Albino-Pimentel, Dussauge \& Shaver (2018); \\
39 & 9 & 6 & 3 & Khanna, Guler \& Nerkar (2018); Hawk \&   Pacheco-de-Almeida (2018); Schepker \& Barker (2018); \\
39 & 8 & 5 & 4 & Yayavaram, Srivastava \& Sarkar (2018); Gandal, Markovich   \& Riordan (2018); Manning, Massini, Peeters \& Lewin (2018); Shi \&   Connelly (2018); \\
39 & 7 & 8 & 4 & Byun, Frake \& Agarwal (2018); Mawdsley \& Somaya (2018);   Alvarez-Garrido \& Guler (2018); Gupta, Mortal \& Guo (2018); \\
39 & 6 & 0 & 0 & - \\
39 & 5 & 9 & 5 & Chen \& Garg (2018); Kaul, Nary \& Singh (2018); Flammer, (2018); Ram\'{i}rez \& Tarzij\'{a}n (2018); Wiersema, Hishimure \& Suzuki (2018); \\
39 & 4 & 8 & 5 & Hawn, Chatterji \& Mitchell (2018); Choudhury \& Haas (2018); Bode \& Singh (2018); Tarakci, Ate\c{s}, Floyd, Ahn \& Wooldridge (2018); Rhee \& Leonardi (2018); \\
39 & 3 & 0 & 0 & - \\
39 & 2 & 6 & 4 & Chen, Kale \& Hoskisson (2018); Choi \& McNamara (2018);   Deichmann \& Jensen (2018); Pek, Oh \& Rivera (2018); \\
39 & 1 & 8 & 3 & Furr \& Kapoor (2018); Vidal \& Mitchell (2018); Jiang,   Xia, Canella \& Xiao (2018); \\
38 & 13 & 8 & 4 & Chem, Qian \& Narayanan (2017); Rabier (2017); Dorobantu   \& Odziemkowska (2017); Li, Yi \& Cui (2017); \\
38 & 12 & 5 & 3 & Lee \& Puranam (2017); Werner (2017); Theeke \& Lee (2017); \\
38 & 11 & 8 & 4 & Carnahan (2017); K\"{o}lbel, Busch \& Jancso (2017); Bos, Faems   \& Noseleit (2017); Li \& Zhou (2017); \\
38 & 10 & 8 & 4 & Moeen (2017); Raffiee (2017); Jiang, Canella, Xia \&   Semadeni (2017); Wei, Ouyang \& Chen (2017); \\
38 & 9 & 8 & 5 & Souder, Zaheer, Sapienza \& Ranucci (2017); Caner, Cohen   \& Pil (2017); Shan, Fu \& Zheng (2017); Wang, Zhao \& Chen (2017); Li,   Xia \& Lin (2017); \\
38 & 8 & 9 & 5 & Zhou \& Wan (2017);    Kulchina (2017); Kim \& Steensma (2017); Steinbach, Holcomb, Holmes,   Devers \& Canella (2017); Makino \& Chan (2017); \\
38 & 7 & 10 & 3 & Armanios, Eesley, Li \& Eisenhardt (2017); Ref \&   Shapira (2017); McCann \& Bahl (2017); \\
38 & 6 & 7 & 3 & Roy \& Cohen (2017); Dowell \& Muthulingam (2017);   Vanacker, Collewaert \& Zahra (2017); \\
38 & 5 & 9 & 5 & Stan \& Puranam (2017); Asgari, Singh \& Mitchell (2017);   Kuusela, Keil \& Maula (2017); Girod \& Whittington (2017); Connelly,   Tihanyi, Ketchen, Carnes \& Ferrier (2017); \\
38 & 4 & 8 & 1 & Silverman \& Ingram (2017); \\
38 & 3 & 10 & 4 & Bermiss, Hallen, McDonald \& Pahnke (2017); Chatterjee (2017); Oh \& Oetzel (2017); Blake \& Moschieri (2017); \\
38 & 2 & 11 & 5 & Flammer \& Luo (2017); Madsen \& Walker (2017); Mackey, Barney \& Dotson (2017);   Fonti, Maoret \& Whitbred (2017); Deb, David \& O'Brien (2017); \\
38 & 1 & 0 & 0 & - \\
37 & 13 & 6 & 2 & Hawn \& Ioannou (2016); Stuart \& Wang (2016); \\
37 & 12 & 7 & 3 & Wang, Zhao \& He (2016); Easley, Decelles \& Lenox (2016);   Wu \& Salomon (2016); \\
37 & 11 & 10 & 7 & Ghosh, Ranganathan \& Rosenkopf (2016); Kalnins (2016);   Chang, Kogut \& Yang (2016); Tsang \& Yamanoi (2016); Massimo, Colombo   \& Shafi (2016); Chadwick, Guthrie \& Xing (2016); Park, Borah \&   Kotha (2016); \\
37 & 10 & 8 & 2 & Husted, Jamali \& Saffar (2016); Van Reenen \& Pennings, (2016); \\
37 & 9 & 6 & 1 & Gomulya \& Boeker (2016); \\
37 & 8 & 10 & 5 & Fonti \& Maoret (2016); Rodr\'{i}guez \& Nietro (2016); Zhu   \& Yoshikawa (2016); Yu, Umashankar \& Rao (2016); Jain (2016); \\
37 & 7 & 12 & 3 & Bennet \& Pierce (2016); Anand, Mulotte \& Ren (2016);   Geng, Yoshikawa \& Colpan (2016); \\
37 & 6 & 8 & 3 & Smith \& Chae (2016); Klingebiel \& Joseph (2016); Karna,   Richter \& Riesenkampf (2016); \\
37 & 5 & 5 & 4 & Roy \& Sarkar (2016); Lungeanu, Stern \& Zajac (2016);   Tyler \& Caner (2016); Brandes, Dharwadkar \& Suh (2016); \\
37 & 4 & 6 & 4 & Adner \& Kapoor (2016); Maslach (2016); Poppo, Zhou \& Li,   2016; Eckhardt (2016); \\
37 & 3 & 9 & 3 & Feldman, Amit \& Villalonga (2016); Pe'er, Vertinsky \&   Keil (2016); Barroso, Giarratana, Reis \& Sorenson (2016); \\
37 & 2 & 8 & 3 & Chen, Crossland \& Huang (2016); Desender, Aguilera, L\'{o}pezpuertas‐Lamy \& Crespi (2016); Kang (2016) \\
37 & 1 & 6 & 2 & Dezs\"{o}, Ross \& Uribe (2016); Ge, Huang \& Png (2016); \\
36 & 13 & 8 & 3 & Joseph \& Gaba (2015); Macher \& Mayo (2015); Zhu \&   Chen (2015); \\
36 & 12 & 7 & 3 & Fuentelsaz, Garrido \& Maicas (2015); Malhotra, Zhu \&   Reus (2015); Chen (2015); \\
36 & 11 & 8 & 5 & Zheng, Singh \& Mitchell (2015); Speckbacher, Neumann \&   Hoffmann (2015); Skilton \& Bernardes (2015); Bermiss \& Murmann (2015);   Fosfuri, Giarratana \& Roca (2015); \\
36 & 10 & 4 & 3 & Kaplan \& Vakili (2015); Chen, Crossland \& Luo (2015);   Ang, Benischke \& Doh (2015); \\
36 & 9 & 7 & 4 & Chittoor, Kale \& Puranam (2015); Chang \& Shim (2015);   Banalieva, Eddleston \& Zellweger (2015); Hashai (2015); \\
36 & 8 & 8 & 4 & Bidwell, Won, Barbulescu \& Mollick (2015); Steensma, Chari   \& Heidl (2015); Durand \& Vergne (2015); Lange, Boivie \& Westphal (2015); \\
36 & 7 & 9 & 6 & Elfenbein \& Knott (2015); Blettner, He, Hu \& Bettis (2015); Arrfelt, Wiseman, McNamara \& Hult (2015); Ioannou \& Serafeim (2015); Wowak, Mannor \& Wowak (2015); Pacheco \& Dean (2015); \\
36 & 6 & 7 & 3 & Kim (2015) (Appendix); Chizema, Liu, Lu \& Gao (2015);   Miller, Xu \& Mehrotra (2015); \\
36 & 5 & 7 & 5 & Bertrand \& Capron (2015); Ganco, Ziedonis \& Agarwal,   2015; Younge, Tong \& Fleming (2015); Damaraju, Barney \& Makhija (2015);   Madsen \& Rodgers (2015); \\
36 & 4 & 6 & 3 & Greve \& Seidel (2015); Harmon, Kim \& Mayer (2015); Tortoriello (2015); \\
36 & 3 & 4 & 3 & Diestre, Rajagopalan \& Dutta (2015); Chadwick, Super \&   Kwon (2015); Kapoor \& Furr (2015); \\
36 & 2 & 6 & 4 & Pacheco-de-Almeida, Hawk \& Yeung (2015); Chown \& Lui (2015); Argyres, Bigelow \& Nickerson (2015); Tong, Reuer, Tyler \& Zhang (2015); \\
36 & 1 & 2 & 1 & Cain, Moore \& Haran (2015);\\* \bottomrule
\end{longtable}

%% file: tables/appendix_orgsci.tex

\begin{longtable}[!]{ccccp{9cm}}
\caption{Organization Science} \\
\toprule
\textbf{Volume} & \textbf{Issue} & \textbf{Total} & \textbf{Count} & \textbf{Articles} \\* \midrule
\endhead
\bottomrule
\endfoot
\endlastfoot
31 & 6 & 7 & 3 & Park \& Zhang (2020); Maoret, Tortoriello \& Iubatti (2020); Assenova (2020); \\
31 & 5 & 6 & 2 & Giarratana \& Santaló (2020); Tasseli, Zappa \& Lomi (2020);                \\
31 & 4 & 5 & 3 & Chown (2020); Boone \& Özcan (2020); Moreira, Klueter \& Tasseli (2020);\\
31 & 3 & 9 & 5 & Claes \& Vissa (2020); Younkin \& Kaskooli (2020);   Tilleman, Russo \& Nelson (2020); Withers, Howard \& Tihanyi (2020); Aharonson, Bort \& Woywode (2020); \\
31 & 2 & 6 & 2 & Chambers \& Baker (2020); Hallen, Cohen \& Bingham (2020); \\
31 & 1 & 5 & 2 & Diestre \& Santaló (2020); Jacqueminet (2020); \\
30 & 6 & 7 & 4 & Rahmandad \& Vakili (2019); Bowers \& Prato (2019); Negro \& Olzak (2019); Rietveld, Schilling \& Bellavitis (2019); \\
30 & 5 & 5 & 3 & Moore, Payne, Filatotchev \& Zajac (2019); Berchicci, Dutt \& Mitchell (2019); Furr (2019); \\
30 & 4 & 8 & 2 & Gaba \& Greve (2019); Bird, Short \& Toffel (2019); \\
30 & 3 & 6 & 2 & Lee (2019); Blevins, Sauerwald, Hoobler \& Robertson (2019); \\
30 & 2 & 8 & 3 & Chatman, Greer, Sherman \& Doerr (2019); Furlan, Galeazzo (2019); Eklund \& Kapoor (2019); \\
30 & 1 & 9 & 4 & Knott \& Turner (2019); Zhang (2019); Rockart \& Wilson (2019); Godart \& Galunic (2019); \\
29 & 6 & 8 & 1 & Keum \& Eggers (2018); \\
29 & 5 & 9 & 4 & Furlotti \& Soda (2018); Berry (2018); Albert (2018); Ertug, Gargiulo, Galunic \& Zou (2018); \\
29 & 4 & 5 & 4 & Zhao, Ishihara, Jennings \& Lounsbury (2018); Hiatt, Carlos \& Sine (2018); James \& Vaaler (2018); Madsen \& Desai (2018); \\
29 & 3 & 7 & 3 & Knoben, Oerlemans, Krijkamp \& Proven (2018); Lawrence (2018); Wry \& Zhao (2018); \\
29 & 2 & 5 & 2 & Maslach, Branzei, Rerup \& Zbaracki (2018); Rietveld \& Eggers (2018); \\
29 & 1 & 7 & 3 & Conti (2018); Durand \& Georgallis (2018); Hsu, Koçak \& Kovács (2018); \\
28 & 6 & 6 & 1 & Greve \& Yue (2017); \\
28 & 5 & 6 & 4 & Ditriadis, Lee, Ramarajan \& Battilana (2017); Tzabbar \&   Margolis (2017); Lee \& Kapoor (2017); Ozmel, Yavuz, Reuer \& Zenger (2017); \\
28 & 4 & 5 & 3 & Merluzzi (2017); Keum \& Kelly (2017); Mannucci (2017); \\
28 & 3 & 8 & 3 & Laursen, Moreire, Reichstein \& Leone (2017); Kapoor \&  Agarwal (2017); Reuer \& Devarakonda (2017); \\
28 & 2 & 7 & 2 & Chattopadhyay \& Choudhury (2017); Bonet \& Salvador (2017); \\
28 & 1 & 8 & 7 & Obloj \& Zenger (2017); Lee \& Meyer-Doyle (2017); Gartenberg \& Wulf (2017); Hoehn-Weiss, Karim \& Lee (2017); McEvily,   Zaheer \& Kamal (2017); Ferguson \& Carnabuci (2017); Eberhart, Eesley   \& Eisenhardt (2017); \\
27 & 6 & 8 & 4 & Quintane \& Carnabuci (2016); Boone \& Özcan (2016); Mollick (2016); Vanacker \& Forbes (2016); \\
27 & 5 & 11 & 7 & Wang, Doucet, Waller, Sanders \& Phillips (2016); Sako,   Chondrakis \& Vaaler (2016); Choi, Kumar \& Zambuto (2016); Souder,   Reilly, Bromiley \& Mitchell (2016); Yang \& Schwartz (2016); Eesleym (2016); Zhang, Marquis \& Qiao (2016); \\
27 & 4 & 8 & 6 & Stenard \& Sauermann (2016); Hahl (2016); Montauti \&   Wezel (2016); Lin (2016); Burbano (2016); Bhaskarabhatla (2016); \\
27 & 3 & 0 & 0 & - \\
27 & 2 & 8 & 5 & Szulanski, Ringov \& Jensen (2016); Zhang \& Gimeno (2016); Kilduff,   Willer \& Anderson (2016); Khessina \& Reis (2016); Eesley, Li \&   Yang (2016); \\
27 & 1 & 8 & 5 & Adams, Fontana \& Malerba (2016); McDonell (2016); Slavova, Fosfuri \& De Castro (2016); Cuypers, Koh \& Wang (2016); Kozhikode (2016); \\
26 & 6 & 9 & 4 & Hasan, Ferguson \& Koning (2015); Carnabuci, Operti \& Kovács (2015); Huang \& Washington (2015); Hiatt, Grandy, \& Lee (2015); \\
26 & 5 & 12 & 3 & Cobb (2015); Srivastava (2015); Verhaal, Khessina \& Dobrev (2015); \\
26 & 4 & 10 & 5 & Sosa, Gargiulo \& Rowles (2015); Marino, Aversa, Mesquita \& Anand (2015); Vidal \& Mitchell (2015); Brands, Menges \&b Kilduff (2015); Kleinbaum, Jordan \& Aufia (2015); \\
26 & 3 & 13 & 7 & Sterling (2015); Dobrajska, Billinger \& Karim (2015); Piazza \& Perretti (2015); O'Reilly, Robinson, Berdahl \& Banki (2015); Lee \& Lounsbury (2015); Jacobides \&Yae (2015); Yang, Li \& Delios, (2015); \\
26 & 2 & 13 & 6 & Williams \& Polman (2015); Rider \& Tan (2015); Casciaro \& Lobo (2015); Tortoriello, McEvily \& Krackhardt (2015); Gambardella, Ganco \& Honoré (2015); Evans, Hendron \& Oldroyd (2015); \\
26 & 1 & 14 & 7 & Jensen \& Kim (2015); Smith \& Hou (2015); Lo \& Kennedy (2015); Almeida, Phene \& Li (2015); Aggarwal \& Wu (2015); Zhou (2015); Lui \& Wezel (2015);\\* \bottomrule

\end{longtable}

%% file: tables/quotes.tex
\begin{longtable}{p{0.15\linewidth} | p{0.85\linewidth}}
    \caption{Examples quotes showing how authors interpret control variable estimates}\label{tab:quotes}\\
    \hline
        Category & Example Quotes \\ \hline
       \multirow{3}{\linewidth}{Control variables behave as expected} 
       
       & \onehalfspacing"In all models, the control variables are generally as expected. Total assets have positive and statistically significant effects on divestment and acquisition frequencies (p = 0.000 in all models), which is expected because larger firms divest and acquire more in absolute terms." (Kuusela, Keil, and Maula, 2017, p. 1111), \emph{Strategic Management Journal} \\ 
       
       & \onehalfspacing"The effects of control variables are as we expected and relatively consistent across models. First, the number of hotels in product segments has the expected effects. As the number of hotels increases, especially in the luxury segment, new units tend to be established farther from rival units (closer to sister units). On the other hand, we observe the opposite effect as the number of upscale hotels increases." (Woo, Cannella \& Mesquita, 2019, p. 1775-1776), \emph{Strategic Management Journal}\\
       
       & \onehalfspacing"As expected, subsidiaries in higher GDP countries were more likely to be supervised locally. Telecommunication positively affected local supervision and tax rate negatively affected local supervision, although these coefficients are not statistically significant. Column (2) adds quality of institutions; it had a significant and positive impact on local supervision." (Zhou, 2015, p. 287), \emph{Organization Science} \\ \hline
       
       \multirow{3}{\linewidth}{Noting interesting control variable estimates} 
        
        & \onehalfspacing"The coefficients of several control variables offered some interesting insights on technology exit. First, the coefficient of Number of competitors was positive ($\beta = 0.35, s.e. = 0.15$) with a p-value of .01. We then computed \(\displaystyle [100[\exp(\beta)-1]\), which gave the percentage change in the hazard associated with a one-unit increase in covariate. The results suggested that if the Number of competitors increased by one, then there was a 41\% increase in the focal technology’s hazard of technology exit." (Chen, Qian, and Narayanan, 2017, p. 2588), \emph{Strategic Management Journal} \\
        
        & \onehalfspacing"A number of results for the control variables are noteworthy. Business line diversification negatively affects the diversification and dissimilarity of technology-sourcing portfolios. The reason may be that business line diversification diverts financial and human resources from investments in technology-sourcing vehicles. We note that the size of the top management team is also negatively related to both diversification and dissimilarity (although not significant for the latter)." (Lungeanu, Stern and Zajac, 2016, p. 865), \emph{Strategic Management Journal} \\
        
        & \onehalfspacing"A few results from Model 1 are worth noting. First, the estimates of the controls appear aligned with our expectations: Population age exhibits a negative and significant effect on growth rates. ... The effects of MU density and MU density2 are consistent with existing evidence on the consequences of density-dependent legitimation and competition on organizational growth rates. C4 market share exhibits a positive and significant effect on MU growth—a result aligned with the current understanding of resource partitioning." (Liu \& Wezel, 2015, p. 301), \emph{Organization Science} \\ \hline

        \multirow{3}{\linewidth}{Confirming prior findings}
        
        & \onehalfspacing"Facility experience has a significantly negative influence on waste change (i.e., the longer a facility has been reporting on a particular chemical, the greater the improvement in operational performance). One year of facility experience improves operational performance by 1.49\%; this translates to a real waste reduction of about 7,300 pounds. This finding is consistent with existing research that links experience and problem-solving outcomes positively while indicating that the facilities in the study have not reached diminishing marginal returns." (Berchicci, Dutt \& Mitchell, 2019, p. 1041), \emph{Organization Science} \\
        
        & \onehalfspacing"we ran the regression only with the control variables and found that inventor productivity, tenure age, and firm concentration in the focal patent class tended to have statistically significant effects and expected signs. The positive effect of patenting productivity was consistent with results reported in prior work (Hoisl 2007, Palomeras and Melero 2010). The negative effect of tenure also was consistent, suggesting that more experienced employees tended to stay (Braguinsky et al. 2012). The positive effect of age was to be expected in our sample of employees with high human capital, as transitioning to entrepreneurship may require extensive prior work experience." (Gambardella, Ganco, and Honoré, 2015, p. 466), \emph{Organization Science} \\
        
        & \onehalfspacing"The sign and direction of control variables align with what has already been documented extensively (Clement, 1999; Clement, Hales, \& Xue, 2011; Hong \& Kubik, 2003; Hong, Kubik, \& Solomon, 2000; Irvine, 2004; Loh \& Mian, 2006)." (Uribe, 2020, p. 1920), \emph{Strategic Management Journal} \\ \hline

        \multirow{3}{\linewidth}{Ex-post rationalizations based on theory} 
        
        & \onehalfspacing"Of the control variables in Models (1)–(3), we see that greater lagged count of posted reviews has a negative effect on dispensaries’ focus on medical use in their identity statements. One possibility for this negative coefficient is that as a dispensary’s customer base increases, it tends to move to broaden its appeal by decreasing its earlier medical-use orientation." (Hsu, Koçak, and Kovács, 2018, p. 184), \emph{Organization Science} \\
        
        & \onehalfspacing"Interestingly, we did not find a relationship between our dependent variable and the passage of a state RPS or state-level incentives. There are a variety of explanations for these results. First, while many states in our sample have passed an RPS, most of these policies were enacted on or after 2004, and have set goals for or after the year 2020. Since our data sample ends in 2009, it is likely that there is insufficient time lapse to observe substantial effects. In addition, provisions behind the PTC might prevent some projects from taking full advantage of state-level incentives and diminish their overall value (Wiser, Bolinger, and Gagliano, 2002)." (Pacheco \& Dean, 2015, p. 1098), \emph{Strategic Management Journal} \\
        
        & \onehalfspacing"Model 1 is the baseline formulation with all the controls. Older banks were more likely to enter dormancy. Perhaps these banks have learned from their experience under previous detrimental policies that were later reversed. Strong recent performance under the detrimental public policy encourages a bank to enter dormancy. This is perhaps a factor of the overconfidence bred by recent good performance. Banks with a higher deposits-to-liability ratio did not prefer dormancy. This indicates that, as expected, a greater reliance on branch banking reduces a bank’s capacity to enter dormancy." (Kozhikode, 2016, p. 199), \emph{Organization Science} \\ \hline
         
\end{longtable}